\documentclass[twocolumn,showpacs,preprintnumbers,amsmath,amssymb]{revtex4}
\usepackage{graphicx}% Include figure files
\usepackage{dcolumn}% Align table columns on decimal point

\begin{document}

%\preprint{APS/123-QED}

\title{Fast intersystem crossing in transition-metal complexes}

\author{Michel van Veenendaal, Jun Chang, and A. J. Fedro}
\affiliation{%
Dept. of Physics, Northern Illinois University,
De Kalb, Illinois 60115\\ 
Advanced Photon Source,
  Argonne National Laboratory, 9700 South Cass Avenue, Argonne,
Illinois 60439}%

%\affiliation{

\date{\today}% It is always \today, today,

\begin{abstract}
The mechanism behind fast intersystem crossing in transition-metal 
complexes is shown to be a result of the dephasing of the photoexcited state
to the phonon continuum of a different state with a significantly 
different transition metal-ligand distance. The coupling is a result
of the spin-orbit interaction causing a change in the local moment.
Recurrence to the initial state 
is prevented by the damping of the phonon oscillation.
The decay time is faster than the oscillation frequency of the 
transition metal-ligand stretch mode, in agreement with experiment.
For energies above the region where the strongest coupling occurs, 
a slower ``leakage''-type decay is observed. If the photoexcited state
is lower in energy than the state it couples to, then there is no decay.
\end{abstract}

\maketitle

{\it Introduction.$-$}
Fast intersystem crossing is an intriguing phenomenon that has puzzled 
many for several decades 
\cite{Decurtins,Hauser,Gutlich,Gawelda,GaweldaPRL,Bressler,Ogawa}. 
In a wide variety of transition-metal
complexes, laser excitation creates a photoinduced excited state 
that decays on the order of tens to hundreds of femtoseconds 
into a state with often a different spin and a significant change in
transition metal-ligand distance. This is generally followed by 
a cascade of intersystem crossings.  
Examples are spin crossover phenomena in divalent iron.
The iron atom is generally surrounded by an 
organic material, such as Fe[(phen)$_2$ (NCS)$_2$] 
or Fe(bpy)$_3$ (bpy=bipyridine) \cite{Decurtins,Hauser, Gutlich}. 
In the ground state, Fe$^{2+}$ is in a low spin state.
 Illumination by light causes a charge 
transfer to the ligands, followed by a cascade of 
intersystem crossings, turning the singlet configuration ($S=0$)
into a high-spin $S=2$ state. 
The high-spin state has a temperature-dependent
 decay time back into the low-spin state that can vary from nanoseconds up to days.
Comparable crossovers have been observed in nickel compounds, where 
the electronic transition is followed by  dissociation 
of the molecule \cite{Chen}. 
These phenomena have been explored since the sixties 
predominantly using optical and M\"ossbauer spectroscopy \cite{Gutlich}. 
 Recent ultrafast x-ray spectroscopic measurements 
 \cite{Gawelda,GaweldaPRL,Bressler} have provided more detailed 
information on the crucial first step that sets up the cascading process
showing that the electronic transition occurs on the order of tens of fs 
\cite{Gawelda} whereas the lattice relaxes on a timescale related to 
the transition metal-ligand stretch mode (100-200 fs) \cite{Bressler}. 

Experimental work has focused on identifying the states and 
relevant times scales of the intersystem crossing. Theory has followed
several approaches. Study of the time-dependence generally relies
on phenomenological rate equations \cite{Hauser,Koshino}. 
Energy level diagrams have been obtained using {\it ab initio} techniques
that use an adiabatic approximation \cite{Suaud,Ordejon}, 
which has the disadvantage that strong coupling between states only occurs for 
very particular energies and transition metal-ligand distances.

This Letter focuses on the important first intersystem crossing and explains
why the system decays so rapidly into states that dramatically differ
in both  spin and lattice parameter with  quantum efficiencies
 close to 100\%.  Several types
of decay depending on the relative energy positions of 
the states are identified.

\begin{figure}[b]
\begin{center}
\includegraphics[angle=0,width=0.65 \linewidth]{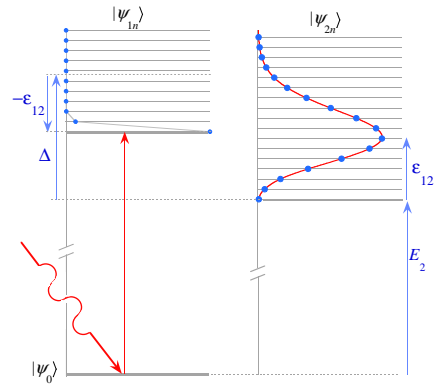}
\caption{\label{eps}(color online) Schematic  of 
fast intersystem crossing. The initial photoexcitation from 
state $|\psi_0\rangle$ into state $|\psi_{1n}\rangle$ occurs with minimal
lattice distortion, i.e., the number of excited phonons $n$
is small. State $|\psi_{1n}\rangle$
couples to state $|\psi_{2n}\rangle$ under the excitation of phonons.
The coupling strength (red curve) creates an effective
phonon density of states.  }
\label{schematic}
\end{center}
\end{figure}

{\it Excitation.$-$} Figure \ref{schematic} gives a schematic diagram
of the fast intersystem crossing. Initially, the system is in the ground 
state $|\psi_0\rangle$. It is then excited into state 
$|\psi_1\rangle$ from which it will decay into state $|\psi_2\rangle$. 
Depending on the change in coupling to the ligands,
several phonons can be created in the excitation process, leading 
to states $|\psi_{1n}\rangle$. 
Although details of the initial excitation can play a role, its primary 
role is to create an excited state higher in energy compared to
states that have a different spin.

{\it Dephasing.$-$}
We now turn our attention to the crucial step of the
 fast intersystem crossing. 
 Transitions between different spin states are accompanied by a change in
lattice constant due to the conversion of $t_{2g}$  into
$e_g$ electrons  that repel the ligands more strongly
(for example,  from LS $t_{2g}^6$ to HS $t_{2g}^4e_g^2$).
This can be described by the Hamiltonian
\begin{eqnarray}
H_0=\sum_{i=1,2} E_i c_i^\dagger c_i+\hbar\omega  a^\dagger a
+\sum_{i=1,2}\sqrt{\varepsilon_i\hbar \omega} n_i (a^\dagger+a),
\end{eqnarray}
where $E_i=E_2+\Delta, E_2$ for the relative positions of states 1 and 2, respectively.
 The third term describes the coupling between the electrons
and the lattice. Classically, this term corresponds to a constant 
force displacing the equilibrium position of the ligand atoms.
Only the relative change in coupling is of importance. 
 Since experimentally one observes a change in spin of $\Delta S=1$,
the coupling,
\begin{eqnarray}
H_1=V(c_1^\dagger c_2+c_2^\dagger c_1),
\end{eqnarray}
 is generally accepted to be due to the spin-orbit coupling.

\begin{figure}[b]
\begin{center}
\includegraphics[angle=0,width=0.8 \linewidth]{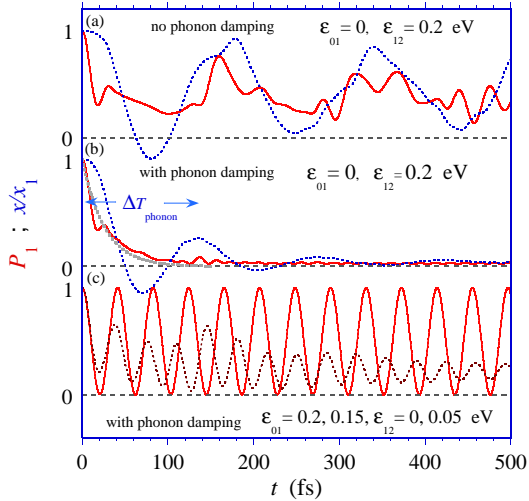}
\caption{\label{eps200}(color online) The probability  $P_1$(red) of finding 
state 1 and the relative change in transition metal-ligand
distance (blue) with 
$x/x_1=1,0$ for the equilibrium positions of 1 and 
2, respectively, as a function of the time $t$ after the photoexcitation.
 (a) No phonon damping; $\varepsilon_{ij}$ 
indicates the change in coupling to the lattice
between states $i$ and $j$. Here the change in lattice
 occurs between states 1 and 2; (b) idem, but with phonon damping;
(c) with phonon damping, but now the strongest change in lattice parameters
occurs in the photoexcitation from 0 to 1 (only $P_1$ is shown). 
The total change is $\varepsilon_{02}=0.2$ eV 
and  $\varepsilon_{01}=0.2$ (solid) and 0.15 (dotted) eV.  }
\end{center}
\end{figure}

We can diagonalize $H_0$  with a displaced-harmonic-oscillator transformation 
${\bar H}_0=e^S H_0e^{-S}$ with $S=\sum_i n_i \sqrt{g_i} (a^\dagger-a)$
with $g_i=\varepsilon_i/\hbar \omega$, giving
\begin{eqnarray}
{\bar H}_0=\sum_i (E_i-\varepsilon_i) c_i^\dagger c_i+\hbar\omega a^\dagger a,
\end{eqnarray}
with eigenstates $|\psi_{in}\rangle$ for  states $i=1,2$ 
and $n$ excited phonon modes. The 
transformed coupling $\bar{H}_1=e^S H_1 e^{-S}$  with
$\bar{V}_{nn'}=\langle \psi_{2n}|\bar{H}_1|\psi_{1n'}\rangle $
 is no longer phonon conserving. 
If the change in transition metal-ligand distance in the photoexcitation 
is small,  the prepared state is close to 
$|\psi_{10}\rangle$, see Fig. \ref{schematic}.
The dominant coupling to $|\psi_{2n}\rangle$
 is then $\bar{V}_{n0}^2=e^{-g}g^n/n!$ with  $\sqrt{g}=\sqrt{g_1}-\sqrt{g_2}$. 
The strongest coupling between $|\psi_{10}\rangle$  and $|\psi_{2n}\rangle$ 
occurs at the maximum of the Poisson distribution 
for which $n\cong g$ at an energy $E_2+g\hbar\omega=E_2+\varepsilon_{12}$. 
In the calculations, we take   $\hbar \omega=30$ meV,
which is a typical value for a transition metal-ligand stretching 
mode \cite{Tuchagues}. The value of  $\varepsilon_{12}=\varepsilon_2-\varepsilon_1=0.2$ eV corresponds
to a typical displacement of a few tenths of an \AA ngstrom \cite{Bressler,Ordejon}. This energy difference also corresponds well to the observed change in wavelength of the emission features in the luminescence spectra \cite{Gawelda}. The spin-orbit coupling parameter is
 $V=50$ meV, which is the atomic value calculated in the Hartree-Fock limit \cite{Cowan}.  Let us first consider
  $\Delta=2\varepsilon_{12}$, when $|\psi_{10}\rangle$ coincides with this
maximum, see Fig. \ref{schematic}.
Figure \ref{eps200}(a)  shows the time-dependence 
of the probability $P_1$ of finding the system in state 1, 
with $P_i=\sum_n |\langle \psi_{in}|\psi(t)\rangle|^2$,
and the displacement of ligands 
($x=\langle \psi(t)|\hat{x}|\psi(t)\rangle$, where $x/x_1$=1, 0 correspond
to the equilibrium positions of states 1 and 2, respectively). In
the first 20 fs, we observe a rapid decrease in $P_1$ 
due to the dephasing of  $|\psi_{10}\rangle$ 
into the phonon states $|\psi_{2n}\rangle$. Although dephasing
is more commonly associated with coupling to a continuum \cite{Fano},
there are sufficient phonon states for a fast dephasing.
However, the finite spacing $\hbar\omega$ between the phonon states
 leads to a recurrence as evidenced by the increase in $P_1$
around 140-160 fs. This recurrence is directly related to the 
oscillation of the ligands, where $\hbar\omega=30$ meV corresponds
to a period of 138 fs. When looking at larger times, we see
a finite value of $P_1$, which does not
correspond to the experimental observation of quantum efficiencies
close to 100\%. In addition, there is a continued oscillation of 
the ligands.

{\it Damping.$-$} The quantum efficiency can be dramatically improved 
by including damping of the phonon oscillations \cite{Kuhn}. 
Here we describe how to include a damping to a bath that 
accounts for damping mainly due to intramolecular energy redistribution
\cite{Stannard,Felker}.
In the absence of the coupling $H_1$ relaxes the
oscillations into their equilibrium states $|\psi_{i0}\rangle$. 
 Since  fast intersystem crossings  are
experimentally known to be  almost temperature independent, we take $T=0$. 
The wavefunctions $|\psi(t)\rangle=\sum_{inb}c_{in}^b(t) |\psi_{in}^b\rangle$,
are extended to include  the possible states $b$ of the bath.
The bath variables are integrated out to obtain the
 time-development of the occupation of state $|\psi_{in}\rangle$,
$P_{in}(t)=\sum_b |\langle \psi_{in}^b|\psi(t)\rangle|^2
=\sum_b |c_{in}^b (t)|^2$.
 For the matrix elements 
of  the coupling between the local states and the bath,
$\langle \psi_{j,n-1}^{b} |H_{0B}|\psi_{in}\rangle 
=\sqrt{n}V_{b}\delta_{ij}$,
we take only the terms linear in $a$ and $a^\dagger$. 
The superscript $b$ in the states $|\psi_{in}^{b}\rangle$ indicates that
an additional excitation with energy $E_b$ has been created with
respect to state with the original bath ($|\psi_{in}\rangle$).
From the Schr\"odinger equation, we find for the change in coefficients
due to presence of the bath
\begin{eqnarray}
\left . \frac{dc_{in}(t)}{dt}\right |_B=
-\frac{i}{\hbar}\sum_{b}\sqrt{n}V_{b} c_{i,n-1}^{b}(t),
\label{ci}
\end{eqnarray} 
(for clarity, we omit the subscript $B$ in the
following equations). We omit processes that increase the number of phonons $n$
via processes involving $a^\dagger$ and the creation
of an excitation in the bath (the rotating-wave
approximation). These processes have 
an energy $\hbar\omega+E_b$ and are less likely than the 
phonon conserving excitation of energy $\hbar\omega-E_b$.
Numerical calculations \cite{MvVadd} show that the probability
of these processes are an order of magnitude smaller for the damping 
time scales under consideration in this paper. 
The coefficients on the right-hand side of Eqn. (\ref{ci}), can be found
from
$ dc_{i,n-1}^{b}(t)/dt=-\frac{i}{\hbar}\sqrt{n}V_{b} c_{in}(t)$.
Integration gives for the change in the norm Eqn. (\ref{ci})
\begin{eqnarray}
&& \frac{d|c_{in}(t)|}{dt}=  -n\sum_{b}
\left (\frac{V_{b}}{\hbar}\right )^2 
\int_0^t dt' e^{-i\omega_{i,n-1}^{b}(t-t')} c_{in}(t'),  \nonumber
\end{eqnarray} 
where $\hbar \omega^b_{in}= E_i-\varepsilon_i+E_b$
is an eigenstate of $\bar{H}_0$ and the bath Hamiltonian.
The integral is greatly simplified when using the common 
approximation that the sum over the bath integral can be replaced
by an integral over an effective density of states. 
Taking a  density of states $\bar{\rho}$ and 
 coupling constant $\bar{V}$
 in the region where it couples strongly, we obtain
\begin{eqnarray}
\sum_{b}\frac{V^2_{b}}{\hbar^2}
e^{-\frac{i}{\hbar}E_{b}(t-t')}
&\cong&
\frac{\pi \bar{\rho}_B \bar{V}^2}{\hbar}\delta(t-t').
\label{impulse}
\end{eqnarray} 
Introducing the lifetime broadening
$\Gamma=\pi \bar{\rho}_B\bar{V}^2/\hbar$, we obtain
$  dc_{in}/dt = -n\Gamma c_{in}(t)$ .
This term shows the decrease in $c_{in}$ with the bath unchanged in the time
step. However, at the same time $c_{in}$ can increase by the decay
of the state with $n+1$ excited phonons. 
By calculating the probability for 
 $c^b_{i,n}$ and using Eqn. (\ref{impulse}), we obtain
\begin{eqnarray}
\frac{dP_{in}(t)}{dt}&=&-2n\Gamma P_{in}(t)+
2(n+1)\Gamma P_{i,n+1}(t)
\label{rate}
\end{eqnarray}
The obtained rate equation can also be obtained heuristically by  taking an 
oscillator frequency with damping $\omega-i\Gamma$ \cite{Bopp}.
The changes in the coefficients due to damping is included
in the Schr\"odinger equation through \cite{MvVadd}
\begin{eqnarray}
i\hbar  \frac{d|\psi(t)\rangle}{dt}
=(\bar{H}_0+\bar{H}_1)|\psi(t)\rangle
+\frac{i\hbar}{2}\frac{d\ln P(t)}{dt} |\psi(t)\rangle,
\end{eqnarray} 
with $P(t)=\sum_{in} P_{in}(t)|\psi_{in}\rangle \langle\psi_{in}|$.

Experimentally, no oscillations of the ligands are observed indicating that the intramolecular vibrational energy distribution occurs within half an oscillation period of the stretching mode. Figure \ref{eps200}(b) shows that the inclusion of $\Gamma^{-1}=30$ fs
causes  a damping of the oscillation within a few periods
(for the rest, using the same parameters as Fig. \ref{eps200}(b)). The damping 
strongly supresses the recurrence of intensity in state 1. The net result
is a very fast intersystem crossing with a close to 100\% quantum
efficiency in good agreement with experiment. The decay time, however, 
is not simply  related to the oscillation period of 138 fs of the phonon, 
but significantly smaller. A good estimate 
of the decay time is obtained by viewing it as a dephasing of 
state $|\psi_{10}\rangle$ into the phonon continuum of 
that $|\psi_{2n}\rangle$. 
The coupling strengths $V_{n0}=e^{-g}g^n/n!$, can also be viewed
 as a constant coupling $V$ to a density of states, see Fig. \ref{schematic},
given by 
$\rho(\epsilon)= e^{-\epsilon} g^\epsilon/[\hbar\omega\Gamma (\epsilon+1)]$,
with $\epsilon=E/\hbar \omega$ and 
where $\Gamma(\epsilon+1)$ is the gamma function that  allows calculation of 
the factorial for non-integer values. The lifetime broadening $\Gamma_d$
associated with the dephasing is then straightforwardly 
obtained \cite{Fano} through 
$\Gamma_d=\pi V^2 \rho(\Delta-\varepsilon)/\hbar$. This gives a decay
time of $(2\Gamma_d)^{-1}\cong 26$ fs. The exponential 
decay $e^{-2\Gamma_d t}$ provides
a satisfactory agreement with the numerical calculation, see the dotted line
in Fig. \ref{eps200}(b). 
Furthermore, this rapid decay does not require the involvement 
of  high-frequency phonon modes (150-200 meV), 
as was suggested earlier since 
the electronic and lattice relaxation occurs on different time scales.

It has also been suggested \cite{Gawelda} that the major change
in transition metal-ligand distance occurs in the photoexcitation.
Figure \ref{eps200}(c) shows the calculation for that situation 
under the assumption that the total change is constant 
($\varepsilon_{02}=0.2$ eV). 
If there is no difference in bond length between states 1 and 2 
($\varepsilon_{12}=0$ and $\varepsilon_{02}=0.2$), 
one observes oscillations in $P_1$. Note that in this limit, phonons 
are created in the photoexcitation. For $\varepsilon_{12}=0$,
there is no phonon continuum but a simple coupling between the two states. 
When introducing a small difference in lattice 
between states 1 and 2 ($\varepsilon_{12}=0.05$, see 
dashed line in Fig. \ref{eps200}(c)),
we do not find a fast decay nor a close to 100\% quantum efficiency.

\begin{figure}[t]
\begin{center}
\includegraphics[angle=0,width=0.9 \linewidth]{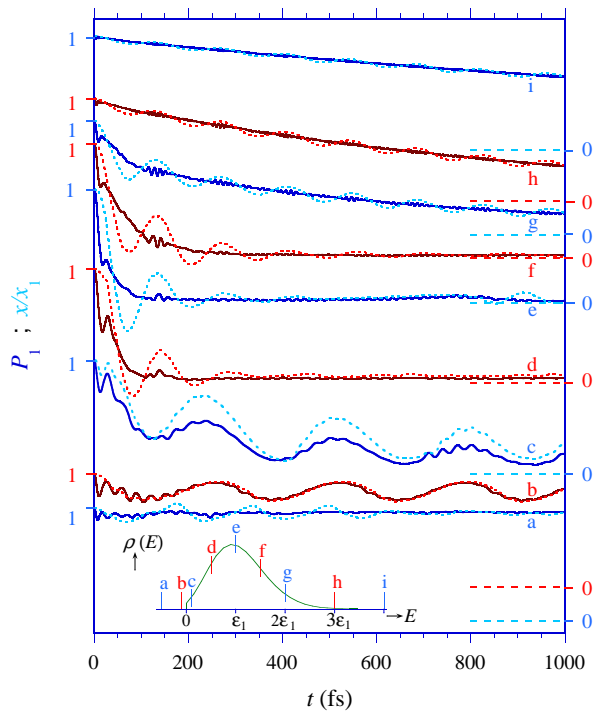}
\caption{\label{delta}(color online)
The probability  $P_1$(alternating dark blue and dark red) of finding 
state 1 and the relative change in transition metal-ligand
distance (alternating dotted blue and red) $x/x_1$ with 
the 1 and 0 values indicated on the left and right axes, respectively,
 in corresponding colors. The dependence as a function of time $t$ is 
given for different positions of $|\psi_{10}\rangle$ with respect to the 
effective phonon density of states $\rho(E)$ of state 2, see 
bottom part of the figure.
  }
\end{center}
\end{figure}
We now turn our attention to the dependence of the relative positions
of states 1 and 2. When using an adiabatic approach,
strong coupling between the different states only occurs for very particular
energies and transition metal-ligand distances. This appears in contradiction
with the prevalence of fast intersystem crossings. We demonstrate that 
the fast decay occurs for a relative wide energy range.
In Fig. \ref{delta}, we show the decay and change in transition metal-ligand
distance for several values of $\Delta$. We take that the initially 
photo-prepared state is $|\psi_{10}\rangle$, i.e. the change in the lattice
upon photoexcitation is small. As described above, the fastest decay 
occurs for  $\Delta=2\varepsilon_{12}$,  see curve e in Fig. \ref{delta}. 
However, also in a region more than $\varepsilon_{12}$ wide
around  this point (see d and f), 
we see the same dephasing with the lifetime broadening
following the effective density of phonon states $\rho(E)$, 
see the lower part of Fig. \ref{delta}. When going
to curve g, there is an initial dephasing followed by a slower 
decrease in $P_1$. However, even at higher energies
(h and i), where the effective phonon density of states $\rho(E)$
is so small that dephasing does not occur, 
the decay occurs on a timescale larger than 
the oscillation period of the phonons. We can understand this
behavior as a ``leakage'' into state 2. 
Since the excitation is into $|\psi_{10}\rangle$, 
phonon damping is irrelevant for state 1. 
The strongest coupling to state 2 occurs at the maximum of the 
effective phonon density of states. For significantly large $\Delta$, we
can treat $\rho(E)$ as a single state and write an effective Green's 
function for state 1
\begin{eqnarray}
G^{-1}(E)= z-\cfrac{V^2}{z-\Delta+2\varepsilon_{12}+i\alpha\Gamma},
\end{eqnarray}
with $z=E-E_2-\Delta+i0^+$.
The phonon damping $\Gamma$ indirectly affects state 1. The effective
state in the Green's function consist of 
 $|\psi_{2n}\rangle$ with $n\cong \varepsilon_{12}/\hbar\omega$.
 Phonon damping causes these state to relax
into $|\psi_{20}\rangle$, which couples very weakly to 
$|\psi_{10}\rangle$. This causes a decrease in $P_1$ with an 
effective decay time 
$\alpha V^2\Gamma/[(\Delta-\varepsilon_{12})^2+(\alpha\Gamma)^2]$. 
The factor $\alpha$ accounts for the fact that there are approximately
$g$ phonons excited.
 For an independent oscillator, the rate equation can 
be solved analytically and is equivalent to $g$ independently decaying 
oscillators with one phonon excited. The time needed to obtain 
a 50\% occupation of the $n=0$ state is $1/\alpha\Gamma$ with 
$\alpha^{-1}=-\ln [1-(1/2 )^{1/g}]$. 

When state $|\psi_{10}\rangle$ is on the low-energy side of the effective 
phonon density of states $\rho(E)$ (a-c in Fig. \ref{delta}),
we see an entirely different behavior. When $\Delta\cong \varepsilon_{12}$,
states $|\psi_{10}\rangle$ 
and $|\psi_{20}\rangle$ are almost degenerate, and one obtains
a strong coupling between them. For, $\Delta > \varepsilon_{12}$, 
state 1 slowly 
decays into state 2. For $\Delta < \varepsilon_{12}$, 
state 1 does not decay into state 2. At first, the asymmetry below and 
above the phonon continuum seems surprising since 
$|\psi_{10}\rangle$ can still
couple to the effective phonon continuum of state 2 and one would expect
leakage. However, now state 2 can easily couple back into the phonon 
continuum $|\psi_{1n}\rangle$  of state 1, 
which is now at a much lower energy (about
$\varepsilon_{12}$ above $|\psi_{10}\rangle$). 
The states $|\psi_{1n}\rangle$ then damp into $|\psi_{10}\rangle$. 

{\it Conclusions}.$-$ An explanation is given for the fast intersystem
crossing in transition-metal systems.  
The large change in lattice between the states
produces an effective phonon continuum leading to a fast dephasing. 
The change in spin is necessary since most parts of the electronic 
Hamiltonian do not produce an intersystem coupling.
Damping of the phonon oscillation by coupling with the environment prevents 
a recurrence of intensity in the initially photoexcited state. 
We demonstrate that an intersystem crossing of tens of fs is compatible
with a lattice relaxation of the order of 100-200 fs.
Different regions of decay can be identified as a function of the 
relative energy 
positions of the states. The fastest decay is an effective dephasing 
and occurs on a smaller timescale than the oscillation period of the phonons 
that are involved in the process. Quantum efficiencies of close to 100\%
are obtained. After obtaining an understanding of the crucial 
initial step leading to a complete spin crossover, future research
should provide a quantum-mechanical understanding of cascading effects 
and finally close the loop with a return to the ground state. 
Additionally, spectroscopic signatures of the process need to be investigated.

{\it Acknowledgments}.$-$ 
This work was  supported by 
 the U.S. Department of Energy (DOE), DE-FG02-03ER46097, and NIU's Institute
for Nanoscience, Engineering, and Technology. Work at
Argonne National Laboratory was supported by the U.S. DOE, 
Office of Science, Office of Basic Energy Sciences, under contract 
DE-AC02-06CH11357.

\end{document}